\documentclass[useAMS,usegraphicx,usenatbib]{mn2e}          

\usepackage{times}

\newcommand{\Msun}{\ensuremath{\,{\rm M}_\odot}}                  
\newcommand{\Rsun}{\ensuremath{\,{\rm R}_\odot}}                  
\newcommand{\Teff}{\ensuremath{T_{\rm eff}}}                      
\newcommand{\logg}{\ensuremath{\log g}}                           
\newcommand{\Mjup}{\ensuremath{\,{\rm M}_{\rm Jup}}}              
\newcommand{\Rjup}{\ensuremath{\,{\rm R}_{\rm Jup}}}              
\newcommand{\Teq}{\ensuremath{T_{\rm eq}^{\,\prime}}}             
\newcommand{\safronov}{\ensuremath{\Theta}}                       
\newcommand{\kms}{\,km\,s$^{-1}$}                                 
\newcommand{\ms}{\,m\,s$^{-1}$}                                   
\newcommand{\mss}{\,m\,s$^{-2}$}                                  
\newcommand{\as}{\ensuremath{^{\prime\prime}}}                    
\newcommand{\FeH}{\ensuremath{\left[\frac{\rm Fe}{\rm H}\right]}} 
\newcommand{\pjup}{\ensuremath{\,\rho_{\rm Jup}}}                 
\newcommand{\psun}{\ensuremath{\,\rho_\odot}}                     
\newcommand{\chir}{\ensuremath{\chi_\nu^{\,2}}}                   
\newcommand{\mc}[1]{\multicolumn{2}{c}{#1}}
\newcommand{\mcc}[1]{\multicolumn{3}{c}{#1}}
\newcommand{\er}[3]{\ensuremath{#1^{+#2}_{-#3}}}
\newcommand{\erc}[3]{\mc{\ensuremath{#1^{+#2}_{-#3}}}}

\newcommand{\ermcc}[5]{\mcc{\ensuremath{{#1\,^{+#2}_{-#3}}\,^{+#4}_{-#5}}}}
\newcommand{\reff}[1]{{#1}}                                   

\title[High-precision defocussed photometry of WASP-17]
      {High-precision photometry by telescope defocussing. IV. Confirmation of the huge radius of WASP-17\,{b}%
      \thanks{Based on data collected by MiNDSTEp with the Danish 1.54\,m telescope at the ESO La Silla Observatory}}

\author[Southworth et al.]
       {John Southworth\,$^{1}$,                                                
        M.\ Dominik\,$^{2}$\thanks{Royal Society University Research Fellow},   
        X.-S.\ Fang\,$^{3}$,                                                    
        K.\ Harps{\o}e\,$^{4,5}$,                                               
        U.\ G.\ J{\o}rgensen\,$^{4,5}$,                                         
        \newauthor
        E.\ Kerins\,$^{6}$,                                                     
        C.\ Liebig\,$^{2}$,                                                     
        L.\ Mancini\,$^{7,8}$,                                                  
        J.\ Skottfelt\,$^{4,5}$,                                                
        D.\ R.\ Anderson\,$^{1}$,                                               
        B.\ Smalley\,$^{1}$,                                                    
        \newauthor
        J.\ Tregloan-Reed\,$^{1}$,                                              
        O.\ Wertz\,$^{9}$,                                                      
        K.\ A.\ Alsubai$^{10}$,                                                 
        V.\ Bozza\,$^{8,11,12}$,                                                
        S.\ Calchi Novati\,$^{8,11}$,                                           
        \newauthor
        S.\ Dreizler\,$^{13}$,                                                  
        S.-H.\ Gu\,$^{3}$,                                                      
        T.\ C.\ Hinse\,$^{14}$,                                                 
        M.\ Hundertmark\,$^{2,13}$,                                             
        J.\ Jessen-Hansen\,$^{15,16}$,                                          
        \newauthor
        N.\ Kains\,$^{17}$,                                                     
        H.\ Kjeldsen,$^{15}$,                                                   
        M.\ N.\ Lund\,$^{15}$,                                                  
        M.\ Lundkvist\,$^{15}$,                                                 
        M.\ Mathiasen\,$^{4}$,                                                  
        M.\ T.\            \newauthor           Penny\,$^{6,18}$,                                               
        S.\ Rahvar\,$^{19,20}$,                                                 
        D.\ Ricci\,$^{9}$,                                                      
        G.\ Scarpetta\,$^{8,11,12}$,                                            
        C.\ Snodgrass\,$^{21}$,                                                 
        J.\ Surdej\,$^{9}$,                                                     
        \\
        $^{1}$\,Astrophysics Group, Keele University, Staffordshire, ST5 5BG, UK \\
        $^{2}$\,SUPA, University of St Andrews, School of Physics \& Astronomy, North Haugh, St Andrews, KY16 9SS, UK \\
        $^{3}$\,National Astronomical Observatories/Yunnan Observatory, Chinese Academy of Sciences, Kunming 650011, China \\
        $^{4}$\,Niels Bohr Institute, University of Copenhagen, Juliane Maries vej 30, 2100 Copenhagen \O, Denmark \\
        $^{5}$\,Centre for Star and Planet Formation, Natural History Museum of Denmark, University of Copenhagen, {\o}ster Voldgade 5-7, 1350 Copenhagen K, Denmark \\
        $^{6}$\,Jodrell Bank Centre for Astrophysics, University of Manchester, Oxford Road, Manchester M13 9PL, UK \\
        $^{7}$\,Max Planck Institute for Astronomy, K\"onigstuhl 17, 69117 Heidelberg, Germany \\
        $^{8}$\,Istituto Internazionale per gli Alti Studi Scientifici (IIASS), 84019 Vietri Sul Mare (SA), Italy \\
        $^{9}$\,Institut d'Astrophysique et de G\'eophysique, Universit\'e de Li\`ege, 4000 Li\`ege, Belgium \\
        $^{10}$\,Qatar Foundation, Doha, Qatar \\
        $^{11}$\,Dipartimento di Fisica ``E. R. Caianiello'', Universit\`a di Salerno, Via Ponte Don Melillo, 84084-Fisciano (SA), Italy \\
        $^{12}$\,Istituto Nazionale di Fisica Nucleare, Sezione di Napoli, Napoli, Italy \\
        $^{13}$\,Institut f\"ur Astrophysik, Georg-August-Universit\"at G\"ottingen, Friedrich-Hund-Platz 1, 37077 G\"ottingen, Germany \\
        $^{14}$\,Korea Astronomy and Space Science Institute, Daejeon, Republic of Korea \\
        $^{15}$\,Stellar Astrophysics Centre (SAC), Department of Physics and Astronomy, Aarhus University, Ny Munkegade 120, DK-8000 Aarhus C, Denmark \\
        $^{16}$\,Nordic Optical Telescope, Apartado 474, E-38700 Santa Cruz de La Palma, Spain \\
        $^{17}$\,European Southern Observatory, Karl-Schwarzschild-Stra{\ss}e 2, 85748 Garching bei M\"unchen, Germany \\
        $^{18}$\,Department of Astronomy, Ohio State University, 140 W. 18th Ave., Columbus, OH 43210, USA \\
        $^{19}$\,Department of Physics, Sharif University of Technology, P.\,O.\,Box 11155-9161 Tehran, Iran \\
        $^{20}$\,Perimeter Institute for Theoretical Physics, 31 Caroline Street North, Waterloo, Ontario N2L 2Y5, Canada \\
        $^{21}$\,Max-Planck-Institute for Solar System Research, Max-Planck Str.\ 2, 37191 Katlenburg-Lindau, Germany \\
        \vspace*{-35pt}}

\begin{document} \maketitle 

\begin{abstract}
We present photometric observations of four transits in the WASP-17 planetary system, obtained using telescope defocussing techniques and with scatters reaching 0.5\,mmag per point. Our revised orbital period is {$4.0 \pm 0.6$\,s longer than previous measurements, a difference of 6.6$\sigma$}, and does not support the published detections of orbital eccentricity in this system. We model the light curves using the {\sc jktebop} code and calculate the physical properties of the system by recourse to five sets of theoretical stellar model predictions. The resulting planetary radius, $R_{\rm b} = 1.932 \pm 0.052 \pm 0.010$\Rjup\ (statistical and systematic errors respectively), provides confirmation that WASP-17\,b is the largest planet currently known. All \reff{fourteen} planets with radii measured to be greater than $1.6$\Rjup\ are found around comparatively hot ($\Teff > 5900$\,K) and massive ($M_{\rm A} > 1.15$\Msun) stars. {Chromospheric activity indicators are available for \reff{eight} of these stars, and all imply a low activity level.} The planets have small or zero orbital eccentricities, so tidal effects struggle to explain their large radii. The observed dearth of large planets around small stars may be natural but could also be due to observational biases against deep transits, if these are mistakenly labelled as false positives and so not followed up.
\end{abstract}



\section{Introduction}                                                                                                              \label{sec:intro}

Ongoing surveys for transiting extrasolar planets (TEPs) have detected an unexpectedly diverse set of objects, such as HD\,80606\,b, a massive planet on an extremely eccentric orbit ($e = 0.9330 \pm 0.0005$; \citealt{Hebrard+10aa}); super-Earths on very short-period orbits like CoRoT-7\,b and 55\,Cnc\,e \citep{Leger+09aa,Winn+11apj}; WASP-18\,b with a mass of 10\Mjup\ and an orbital period of 0.94\,d \citep{Hellier+09nat}; WASP-33\,b, a very hot planet revolving around a metallic-lined pulsating A\,star \citep{Collier+11mn}; a system of six planets transiting the star Kepler-11 \citep{Lissauer+11nat}; and a planet orbiting the eclipsing binary star system Kepler-16 \citep{Doyle+11sci}.

Among these objects, WASP-17\,b stands out as both the largest known planet and the first found to follow a retrograde orbit \citep[][hereafter A10]{Anderson+10apj}. However, the reliability of its radius measurement was questionable for two reasons. Firstly, it rested primarily on a single high-quality transit light curve, whereas it is widely appreciated that correlated noise can afflict individual light curves whilst remaining undetectable in isolation \citep[e.g.][]{Gillon+07aa3,Adams+11apj2}. Correlated noise is clearly visible in the residuals of the best-fitting model for this transit (fig.\,2 in A10). Secondly, the orbital eccentricity, $e$, was poorly constrained, and this uncertainty in the orbital velocity of the planet has major implications for the interpretation of the transit light curves.

The WASP-17 discovery paper (A10) presented three measurements of the planetary radius, $R_{\rm b}$, based on models with different assumptions. The preferred model (Case 1) was a straightforward fit to the available data, yielding $R_{\rm b} = \er{1.74}{0.26}{0.23}$\Rjup\ and $e = \er{0.129}{0.106}{0.068}$. The test of \citet{LucySweeney71aj}, which accounts for the fact that a measured eccentricity is a biased estimator of the true value, indicates a probability of only 83\% that this eccentricity is significant. Case 2 incorporated a Bayesian prior on the stellar mass and radius to encourage them towards a solution appropriate for a main-sequence star, and resulted in $R_{\rm b} = 1.51 \pm 0.10$\Rjup\ and $e = \er{0.237}{0.068}{0.069}$. The third and final model, Case 3, did not use the main-sequence prior but instead enforced $e=0$, and yielded $R_{\rm b} = 1.97 \pm 0.10$\Rjup. The measured size of the planet is clearly very sensitive to the treatment of orbital eccentricity.

\citet{Triaud+10aa} and \citet{Bayliss+10apj} subsequently confirmed the provisional finding that WASP-17\,b has a retrograde orbit, from radial velocity observations obtained during transit. \citet{Triaud+10aa} also obtained improved spectroscopic parameters for the host star. They assumed $e=0$ and found $R_{\rm b} = \er{1.986}{0.089}{0.074}$\Rjup. \citet{Wood+12mn} have detected sodium in the atmosphere of WASP-17\,b, using \'echelle spectroscopy obtained outside and during transit.

\citet[][hereafter A11]{Anderson+11mn} used an alternative method to constrain the orbital shape of the WASP-17 system: measurements of the time of occultation of the planet by the star in infrared light curves obtained by the {\it Spitzer} satellite. To first order, the orbital phase of secondary eclipse (occultation) depends on the product $e\cos\omega$ where $\omega$ is the longitude of periastron \citep{Kopal59book}. A11 obtained $e\cos\omega = 0.00352 \pm 0.00075$ and $e = \er{0.028}{0.015}{0.018}$, finding $e\cos\omega$ to be {significantly different from zero} at the 4.8$\sigma$ level. Inclusion of the {\it Spitzer} data, alongside existing observations, led to the measurement $R_{\rm b} = 1.991 \pm 0.081$\Rjup. This was achieved without making assumptions about $e$ or $\omega$, so represents the first clear demonstration that WASP-17\,b is the largest planet {with a known radius}. One remaining concern was that the orbital ephemeris of the system had to be extrapolated to the times of the {\it Spitzer} observations, potentially compromising the measurement of the phase of mid-occultation and therefore $e\cos\omega$.

In this work we present new photometry of three complete transits of WASP-17\,b, obtained using telescope-defocussing techniques. These lead to a refinement of the orbital ephemeris, shedding new light on the possibility of orbital eccentricity in this system. They also allow a new set of physical properties of the system to be obtained, which are more precise and no longer dependent on the quality of a single follow-up light curve. We use these data to confirm the standing of WASP-17\,b as the largest known planet, at $R_{\rm b} = 1.932 \pm 0.053$\Rjup.


\section{Observations and data reduction}

\begin{table*} \centering
\caption{\label{tab:obslog} Log of the observations presented in this work. $N_{\rm obs}$ is the number
of observations and `Moon illum.' is the fractional illumination of the Moon at the midpoint of the transit.}
\begin{tabular}{llccccccccc} \hline
Transit & Date of   & Start time & End time  &$N_{\rm obs}$ & Exposure & Filter & Airmass & Moon & Aperture   & Scatter \\
        & first obs &    (UT)    &   (UT)    &              & time (s) &        &         &illum.& radii (px) & (mmag)  \\
\hline
1 & 2011 04 28 & 03:48 & 10:10 & 128 & 120     & $R$ & 1.19 $\to$ 1.00 $\to$ 1.57 & 0.212 & 32, 46, 65 & 0.560 \\ 
2 & 2011 05 28 & 00:58 & 03:55 &  63 & 120     & $R$ & 1.31 $\to$ 1.01            & 0.199 & 28, 38, 58 & 0.762 \\ 
3 & 2011 06 11 & 23:51 & 06:51 & 152 & 100     & $R$ & 1.46 $\to$ 1.00 $\to$ 1.48 & 0.826 & 30, 42, 60 & 0.528 \\ 
4 & 2011 06 26 & 23:32 & 05:54 & 152 & 100     & $R$ & 1.26 $\to$ 1.00 $\to$ 1.45 & 0.184 & 30, 40, 60 & 0.475 \\ 
\hline \end{tabular} \end{table*}

\begin{figure} \includegraphics[width=0.48\textwidth,angle=0]{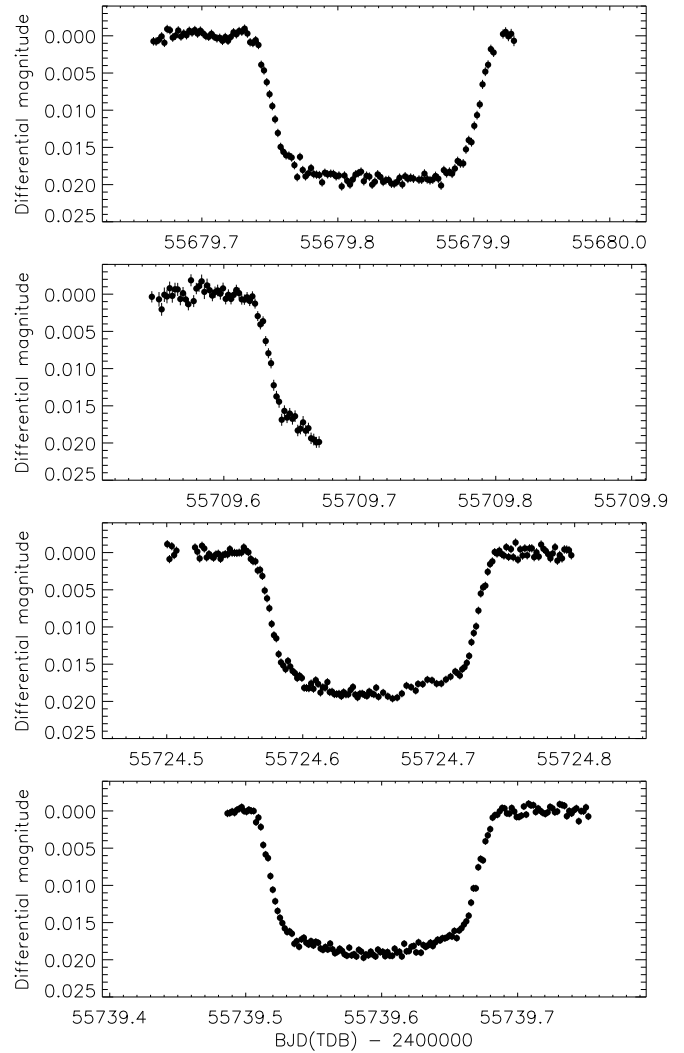}
\caption{\label{fig:plotlc} Light curves of WASP-17. The error bars have been
scaled to give $\chi^2_{\ \nu} = 1.0$ for each night, and in some cases are
smaller than the symbol sizes.} \end{figure}

\begin{table} \centering \caption{\label{tab:lc} Excerpts of the light
curve of WASP-17. The full dataset will be made available at the CDS.}
\begin{tabular}{lrr} \hline
BJD(TDB) & Diff.\ mag. & Uncertainty \\
\hline
55679.66520 &  0.000732 & 0.000572 \\
55679.93036 &  0.000675 & 0.000721 \\[2pt]
55709.54791 &  0.000352 & 0.000761 \\
55709.67092 &  0.019863 & 0.000766 \\[2pt]
55724.50116 & -0.001133 & 0.000560 \\
55724.79836 &  0.000359 & 0.000568 \\[2pt]
55739.48740 &  0.000324 & 0.000508 \\
55739.75282 &  0.000740 & 0.000499 \\
\hline \end{tabular} \end{table}

We observed four transits of WASP-17 through a Bessell $R$ filter, in the 2011 observing season, using the 1.54\,m Danish Telescope%
\footnote{For information on the 1.54\,m Danish Telescope and DFOSC see:
{\scriptsize\tt http://www.eso.org/sci/facilities/lasilla/telescopes/d1p5/}}
at ESO La Silla, Chile (Table\,\ref{tab:obslog}).
Our approach was to defocus the telescope and use relatively long exposure times of 100--120\,s \citep[see][]{Me+09mn,Me+09mn2}. This technique results in a higher observing efficiency, as less time is spent on reading out the CCD, and therefore lower Poisson and scintillation noise. It also greatly decreases flat-fielding noise as several thousand pixels are contained in each point spread function (PSF), and makes the results insensitive to any changes in the seeing during an observing sequence. We autoguided throughout each sequence in order to keep the PSFs on the same CCD pixels, which reduces any remaining susceptibility to flat-fielding noise. The second of the observing sequences suffered from clouds from shortly before the midpoint to after the end of the transit, and we were not able to obtain reliable photometry from the affected data.


Several images were taken with the telescopes properly focussed, in order to check for faint stars within the defocussed PSF of WASP-17. The closest star we detected is 6.9\,mag fainter and separated by 69 pixels (27\as) from the position of WASP-17. Light from this star did not contaminate any of our observations.

Data reduction was performed as in previous papers \citep{Me+09mn,Me+09mn2}, using a pipeline written in {\sc idl}\footnote{The acronym {\sc idl} stands for Interactive Data Language and is a trademark of ITT Visual Information Solutions. For further details see {\tt http://www.ittvis.com/ProductServices/IDL.aspx}.} and calling the {\sc daophot} package \citep{Stetson87pasp} to perform aperture photometry with the {\sc aper}\footnote{{\sc aper} is part of the {\sc astrolib} subroutine library distributed by NASA. For further details see {\tt http://idlastro.gsfc.nasa.gov/}.} routine. The apertures were placed by hand (i.e., mouse-click) and were shifted to follow the positions of the PSFs by cross-correlating each image against a reference image. We tried a wide range of aperture sizes and retained those which gave photometry with the lowest scatter compared to a fitted model. In line with previous experience, we find that the shape of the light curve is very insensitive to the aperture sizes.

Differential photometry was obtained against an optimal ensemble formed from between two and four comparison stars. Simultaneously to optimisation of the comparison star weights, we fitted low-order polynomials to the out-of-transit data in order to normalise them to unit flux. A first-order polynomial (straight line) was used when possible -- such a function is preferable as it is incapable of modifying the shape of the transit -- but a second-order polynomial was needed for the third transit to cope with slow variations induced by the changing airmass.

The final light curves have scatters in the region 0.5\,mmag per point, which is close to the best achieved at this (or any other) 1.5\,m telescope \citep{Me+09apj,Me+10mn}. They are shown individually in Fig.\,\ref{fig:plotlc} and tabulated in Table\,\ref{tab:lc}.


\section{Light curve analysis}                                                                                                         \label{sec:lc}

The analysis of our light curves was performed identically to the {\it Homogeneous Studies} approach established by the first author. Full details can be found in  \citet{Me08mn,Me09mn,Me10mn,Me11mn}. Here we restrict ourselves to a summary of the analysis steps.

The light curves were modelled using the {\sc jktebop}\footnote{{\sc jktebop} is written in {\sc fortran77} and the source code is available at {\tt http://www.astro.keele.ac.uk/jkt/codes/jktebop.html}} code \citep{Me++04mn,Me++04mn2}. The star and planet are represented by biaxial spheroids, and their shape governed by the mass ratio. We adopted the value 0.0004, but our results are extremely insensitive to this number. The salient parameters of the model are the fractional radii of the star and planet (i.e.\ absolute radii divided by semimajor axis) $r_{\rm A}$ and $r_{\rm b}$, and the orbital inclination, $i$. The fractional radii were parameterised by their sum and ratio:
$$ r_{\rm A} + r_{\rm b} \qquad \qquad k = \frac{r_{\rm b}}{r_{\rm A}} = \frac{R_{\rm b}}{R_{\rm A}} $$
as the latter are less strongly correlated.

\subsection{Orbital period determination}                                                                                        \label{sec:lc:porb}

\begin{table} \begin{center}
\caption{\label{tab:minima} Times of minimum light of WASP-17
and their residuals versus the ephemeris derived in this work.
\newline {\bf References:}
(1) A10;
(2) This work.}
\begin{tabular}{l@{\,$\pm$\,}l r r l} \hline
\multicolumn{2}{l}{Time of minimum}   & Cycle  & Residual & Reference \\
\multicolumn{2}{l}{(HJD $-$ 2400000)} & no.    & (HJD)    &           \\
\hline
53890.54887 & 0.00430 &  -188.0 &  0.01843 & 1 \\   
53905.48227 & 0.00380 &  -184.0 &  0.00989 & 1 \\   
53920.42277 & 0.00250 &  -180.0 &  0.00846 & 1 \\   
53965.23767 & 0.00350 &  -168.0 & -0.00246 & 1 \\   
54200.57117 & 0.00310 &  -105.0 & -0.00449 & 1 \\   
54215.52227 & 0.00190 &  -101.0 &  0.00468 & 1 \\   
54271.55737 & 0.00280 &   -86.0 &  0.00751 & 1 \\   
54286.49447 & 0.00580 &   -82.0 &  0.00267 & 1 \\   
54301.45167 & 0.00570 &   -78.0 &  0.01793 & 1 \\   
54331.32347 & 0.00650 &   -70.0 &  0.00585 & 1 \\   
54555.43697 & 0.00440 &   -10.0 & -0.00972 & 1 \\   
54566.65087 & 0.00580 &    -7.0 & -0.00227 & 1 \\   
54592.80046 & 0.00038 &     0.0 & -0.00107 & 1 \\   
55679.82838 & 0.00046 &   291.0 &  0.00084 & 2 \\   
55724.65322 & 0.00056 &   303.0 & -0.00013 & 2 \\   
55739.59522 & 0.00028 &   307.0 & -0.00007 & 2 \\   
\hline \end{tabular} \end{center} \end{table}

\begin{figure*} \includegraphics[width=\textwidth,angle=0]{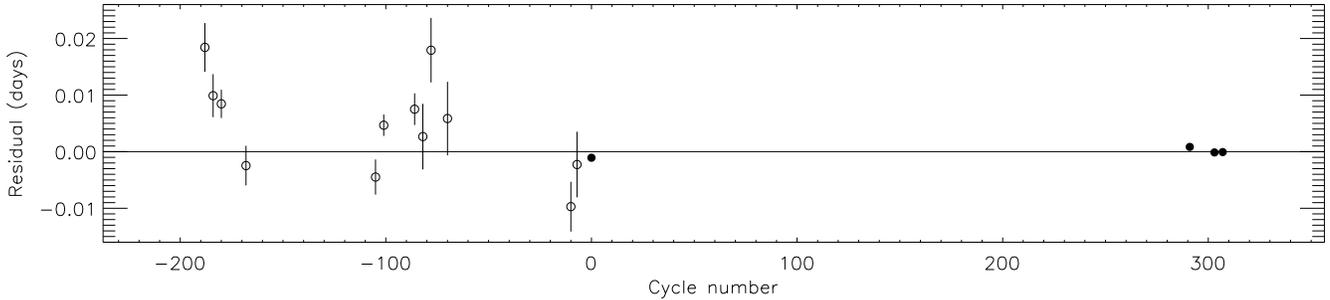}
\caption{\label{fig:minima} Plot of the residuals of the timings of
mid-transit of WASP-17 versus a linear ephemeris. Some error bars
are smaller than the symbol sizes. Timings obtained from SuperWASP
data are plotted using open circles, and other timings (Danish and
Euler telescopes) are plotted with filled circles.} \end{figure*}

Our first step was to obtain a refined orbital ephemeris. Our own four datasets were fitted individually and their errorbars rescaled to give $\chi^2_\nu = 1.0$ with respect to the best-fitting model. This step is necessary because the uncertainties from our data reduction pipeline (specifically the {\sc aper} algorithm) tend to be underestimated. We then re-derived the times of mid-transit for the three datasets which covered complete transits. Monte Carlo simulations were used to assess the uncertainties of the measurements, and the resulting errorbars were doubled based on previous experience \citep{Me+12mn,Me+12mn2}.

We augmented our three times of mid-transit with 13 measurements from A10, of which 12 are from observations with SuperWASP \citep{Pollacco+06pasp} and one is from their follow-up Euler light curve (Table\,\ref{tab:minima}). Taking the time of this follow-up dataset as the reference epoch, we find the ephemeris:
$$ T_0 = {\rm BJD(TDB)} \,\, 2\,454\,592.80154 (50) \, + \, 3.7354845 (19) \times E $$
where $E$ represents the cycle count with respect to the reference epoch. The reduced $\chi^2$ of the fit to the timings is rather large at $\chir = 2.37$, and this has been accounted for in the errorbars above. The dominant contribution to this $\chir$ arises from the SuperWASP timings, which confirms the caveat from A10 that these may have optimistic errorbars.

A plot of the residuals of the fit (Fig.\,\ref{fig:minima}) at first sight suggests the possibility of transit timing variations as might arise from the light-time effect induced by a body on a wider orbit around WASP-17\,Ab, {but a periodogram of the residuals from Table\,\ref{tab:minima} does not show any peaks above the noise level.} We therefore proceeded under the reasonable assumption that the orbital period is constant.

\subsection{Orbital eccentricity}                                                                                       \label{sec:lc:e}

As emphasised in Sect.\,\ref{sec:intro}, the possibility of an eccentric orbit is an important consideration for WASP-17. The radial velocity measurements of the star do not strongly constrain eccentricity; the {radial velocity curve of the star has an amplitude not much greater than the size of the errorbars on the individual measurements}. The observed shape of the transit is not useful because it covers only a very small phase interval (an essentially ubiquitous situation for TEPs; see \citealt{Kipping08mn}). The only precise constraint on orbital shape was obtained by A11, from two occultations observed using {\it Spitzer}. They found the phase of mid-occultation to be $0.50224 \pm 0.00050$, allowing a detection of {a non-zero} $e\cos\omega$ at the 4.8$\sigma$ level.

Our revised period is {$4.0 \pm 0.6$\,s larger than that found by A11, a difference of 6.6$\sigma$}, which affects the phase of mid-occultation. The actual occultation times are not given by A11, but an effective time can be inferred from the dates of the observations and the orbital ephemerides utilised. We performed this calculation and then converted the result back into orbital phase using our new ephemeris. This procedure incorporates the necessary conversion from the UTC to the TDB timescales. We found the phase of occultation to be $0.50066$, which is consistent with phase 0.5 at about the 1$\sigma$ level. The {\it Spitzer} results can no longer be taken as evidence of an eccentric orbit in the WASP-17 system. This emphasises the importance of accompanying occultation measurements with transits, in order to avoid uncertainties in propagating ephemerides from different observing seasons.

To confirm this result, we obtained a time measurement which represents the actual times of the {\it Spitzer} occultations by repeating the analysis by A11. We found $2454949.5422 \pm 0.0016$ on the HJD(UTC) timescale. After converting to the TDB timescale this equates to the phase $0.50059 \pm 0.00043$, which is equivalent to an $e\cos\omega$ of only $0.00093 \pm 0.00068$. This differs from zero at the 1.4$\sigma$ level, which we do not regard as convincing evidence of orbital eccentricity (see also \citealt{Anderson+12mn}). Further observations with Warm {\it Spitzer} would be useful in confirming the phase of mid-occultation of WASP-17\,b.

\subsection{Light curve modelling}                                                                                                  \label{sec:lc:lc}

\begin{table*} \centering \caption{\label{tab:lcfit} Parameters of the fit to the light curves of WASP-17 from the
{\sc jktebop} analysis (top lines). The parameters adopted as final are given in bold. Alternative parameters with
various constraints on orbital eccentricity and orientation are included, labelled with the $e\cos\omega$ value adopted.
The parameters found by other studies are shown in the lowest part of the table. Quantities without quoted uncertainties
were not given by those authors, but have been calculated from other parameters which were. $e\cos\omega$ and
$e\sin\omega$ values are given to show explicitly the measurements or assumptions relevant to each analysis.}
\setlength{\tabcolsep}{3pt}
\begin{tabular}{l r@{\,$\pm$\,}l r@{\,$\pm$\,}l r@{\,$\pm$\,}l r@{\,$\pm$\,}l r@{\,$\pm$\,}l r@{\,$\pm$\,}l r@{\,$\pm$\,}l}
\hline
Source & \mc{$e\cos\omega$} & \mc{$e\sin\omega$} & \mc{$r_{\rm A}+r_{\rm b}$} & \mc{$k$} & \mc{$i$ ($^\circ$)} & \mc{$r_{\rm A}$} & \mc{$r_{\rm b}$} \\
\hline
Danish data & \mc{0.0 assumed}  & \mc{0.0 assumed} & 0.1616 & 0.0021 & 0.1255 & 0.0007 & 86.71 & 0.30 & 0.1436 & 0.0018 & 0.01802 & 0.00030 \\
Euler data  & \mc{0.0 assumed}  & \mc{0.0 assumed} & 0.1744 & 0.0080 & 0.1322 & 0.0012 & 85.46 & 0.83 & 0.1540 & 0.0069 & 0.0204  & 0.0011  \\
Danish data &   0.036 & 0.033   & $-$0.10  & 0.13  & 0.180  & 0.023  & 0.1254 & 0.0007 & 85.9  & 1.1  & 0.160  & 0.021  & 0.0200  & 0.0026  \\
Danish data & 0.00352 & 0.00075 & $-$0.027 & 0.019 & 0.1576 & 0.0040 & 0.1254 & 0.0007 & 86.87 & 0.30 & 0.1400 & 0.0035 & 0.01757 & 0.00047 \\
Danish data & 0.00093 & 0.00068 & $-$0.027 & 0.019 & 0.1591 & 0.0045 & 0.1255 & 0.0007 & 86.81 & 0.32 & 0.1414 & 0.0040 & 0.01774 & 0.00054 \\
\hline
Adopted solution & \mc{0.0 assumed} & \mc{0.0 assumed} & 0.1616 & 0.0021 & 0.1255 & 0.0007 & {\bf 86.71} & {\bf 0.30} & {\bf 0.1436} & {\bf 0.0018} & {\bf 0.01802} & {\bf 0.00030} \\
\hline
A10 (Case 1) & \erc{0.036}{0.034}{0.031} & $-$0.10 & 0.13 & \mc{0.1446} & \erc{0.1293}{0.0011}{0.0014} & \erc{87.8}{2.0}{1.0} & \mc{0.1281} & \mc{0.01658} \\
A10 (Case 2) & \erc{0.034}{0.025}{0.024} & \erc{-0.233}{0.071}{0.070} & \mc{0.1275} & \erc{0.1294}{0.0010}{0.0011} & \erc{88.16}{0.58}{0.45} & \mc{0.1129} & \mc{0.01459} \\
A10 (Case 3) & \mc{0.0 assumed} & \mc{0.0 assumed} & \mc{0.1622} & \erc{0.1295}{0.0010}{0.0010} & \erc{87.8}{2.0}{1.0} & \mc{0.1436} & \mc{0.01855} \\
\citet{Triaud+10aa}&\mc{0.0 assumed}&\mc{0.0 assumed}&\mc{0.1657}&\erc{0.12929}{0.00077}{0.00061}&\erc{86.63}{0.39}{0.45}&\erc{0.1467}{0.0033}{0.0025}&\erc{0.01897}{0.00051}{0.00040}\\
A11 & \erc{0.00352}{0.00076}{0.00073} & \erc{-0.027}{0.019}{0.015} & \mc{0.1605} & 0.1302 & 0.0010 & \erc{86.83}{0.68}{0.56} & \mc{0.1420} & \mc{0.01847} \\
\hline \end{tabular} \end{table*}

\begin{figure} \includegraphics[width=0.48\textwidth,angle=0]{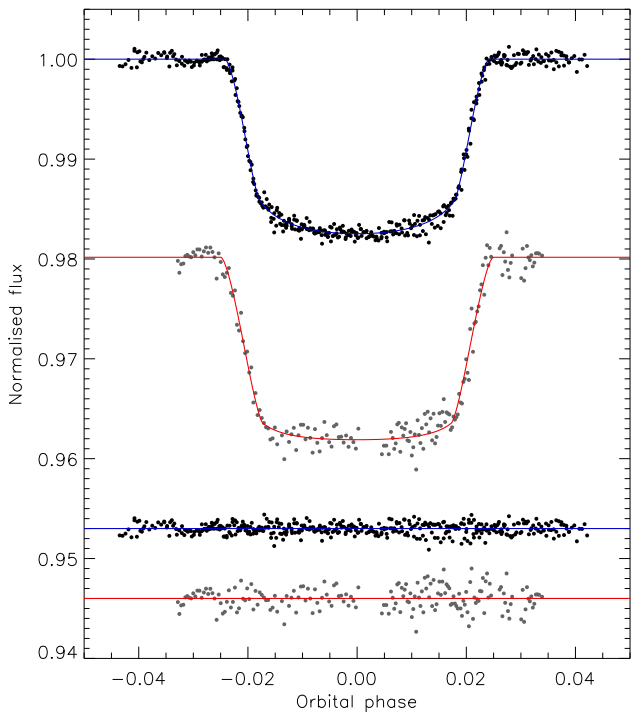}
\caption{\label{fig:lcfit} Phased light curves of WASP-17 from the Danish
telescope (upper) and the Euler telescope (lower), compared to the best fits
found using {\sc jktebop} and the quadratic LD law. The residuals of the fits
are plotted at the bottom of the figure, offset from zero.} \end{figure}

We modelled the three complete transits from the Danish Telescope simultaneously using the {\sc jktebop} code. The partially-observed transit provides a confirmation of the transit depth, but is less reliable than the other ones and has little effect on the solution, so was not included in further analysis. The \chir\ of the fit to the three light curves is $1.22$, which indicates that they don't agree completely on the transit shape. Such a situation may arise from astrophysical effects such as starspot activity {(which can cause changes in the transit depth without altering the shape)}, instrumental effects such as correlated noise in the photometry, and analysis effects such as imperfect transit normalisation. Importantly, our possession of three independent transits mitigates against all three eventualities, making the resulting solutions much more reliable than those based on a single transit. {The excess \chir\ causes larger errorbars to be obtained for both error anlysis algorithms (see below) so is accounted for in our final results.}

Light curve models were obtained using each of five limb darkening (LD) laws, {of which four are biparametric} (see \citealt{Me08mn}). The LD coefficients were either fixed at theoretically predicted values\footnote{Theoretical limb darkening coefficients were obtained by bilinear interpolation in stellar \Teff\ and \logg\ using the {\sc jktld} code available from {\tt http://www.astro.keele.ac.uk/jkt/codes/jktld.html}} or included as fitted parameters. We found that fitting for one LD coefficient provided a significant improvement on fixing both to their theoretical counterparts, but that fitting for both led to ill-conditioned models with no further improvement in the quality of fit. We therefore adopted the fits with the linear LD coefficient fitted and the nonlinear LD coefficient set to its theoretical value but perturbed by $\pm$0.05 on a flat distribution during the error analyses (corresponding to case `LD-fit/fix' in the nomenclature of \citealt{Me11mn}). This does not cause a significant dependence on stellar theory because the two LD coefficients are very strongly correlated \citep{Me++07aa}. {The results for the linear LD law were not used as linear LD is known to be a poor representation of reality.}

Errorbars for the fitted parameters were obtained in two ways: from 1000 Monte Carlo simulations for each solution, and via a residual-permutation algorithm. We found that the residual-permutation method returned larger uncertainties for $k$ but not for other parameters, indicating that red noise becomes important when measuring the transit depth. The final parameter values are the unweighted mean of those from the solutions involving the four two-parameter LD laws. Their errorbars are the larger of the Monte-Carlo or residual-permutation alternatives, with an extra contribution to account for variations between solutions with the different LD laws.

\begin{table*} \centering \caption{\label{tab:model} Derived physical properties of the WASP-17
system. The upper part of the table contains the individual results from this work; in each case
$g_{\rm b} = 3.16 \pm 0.20$\mss, $\rho_{\rm A} = 0.324 \pm 0.012$\psun\ and $\Teq = 1755 \pm 28$\,K.}
\begin{tabular}{l r@{\,$\pm$\,}l r@{\,$\pm$\,}l r@{\,$\pm$\,}l r@{\,$\pm$\,}l r@{\,$\pm$\,}l r@{\,$\pm$\,}l}
\hline
\ & \mc{(dEB constraint)} & \mc{({\it Claret} models)} & \mc{({\it Y$^2$} models)} & \mc{({\it Teramo} models)} & \mc{({\it VRSS} models)} & \mc{({\it DSEP} models)} \\
\hline
$K_{\rm b}$     (\kms) & 148.38  &   3.72  & 148.71  &   2.88  & 148.85  &   1.78  & 149.78  &   0.41  & 148.52  &   1.49  & 145.89  &   2.26  \\
$M_{\rm A}$    (\Msun) & 1.272   & 0.096   & 1.280   & 0.076   & 1.284   & 0.046   & 1.308   & 0.011   & 1.275   & 0.038   & 1.209   & 0.056   \\
$R_{\rm A}$    (\Rsun) & 1.577   & 0.045   & 1.580   & 0.040   & 1.582   & 0.029   & 1.592   & 0.022   & 1.578   & 0.027   & 1.550   & 0.033   \\
$\log g_{\rm A}$ (cgs) & 4.147   & 0.015   & 4.148   & 0.014   & 4.148   & 0.011   & 4.151   & 0.011   & 4.148   & 0.011   & 4.140   & 0.012   \\[2pt]
$M_{\rm b}$    (\Mjup) & 0.473   & 0.035   & 0.475   & 0.033   & 0.476   & 0.029   & 0.482   & 0.027   & 0.474   & 0.028   & 0.457   & 0.029   \\
$R_{\rm b}$    (\Rjup) & 1.925   & 0.058   & 1.929   & 0.052   & 1.931   & 0.040   & 1.943   & 0.033   & 1.927   & 0.038   & 1.893   & 0.043   \\
$\rho_{\rm b}$ (\pjup) & 0.0620  & 0.0049  & 0.0619  & 0.0048  & 0.0618  & 0.0047  & 0.0614  & 0.0046  & 0.0620  & 0.0046  & 0.0631  & 0.0048  \\
\safronov\             & 0.0197  & 0.0012  & 0.0197  & 0.0012  & 0.0196  & 0.0012  & 0.0195  & 0.0011  & 0.0197  & 0.0011  & 0.0200  & 0.0012  \\[2pt]
$a$               (AU) & 0.05105 & 0.00128 & 0.05116 & 0.00099 & 0.05121 & 0.00061 & 0.05153 & 0.00014 & 0.05110 & 0.00051 & 0.05019 & 0.00078 \\
Age              (Gyr) &       \mc{ }      &\erc{2.9}{0.3}{1.0}&\erc{2.7}{0.4}{0.4}&\erc{2.1}{0.0}{0.1}&\erc{2.5}{0.2}{1.2}&\erc{3.3}{0.6}{0.5}\\
\hline \end{tabular} \end{table*}

\begin{table*} \centering \caption{\label{tab:fmodel} Final physical properties of the WASP-17
system (with statistical and systematic errorbars) compared to published measurements. Eccentricity
is included to illustrate the difference approaches taken to obtain each set of results.}
\begin{tabular}{l r@{\,$\pm$\,}c@{\,$\pm$\,}l r@{\,$\pm$\,}l r@{\,$\pm$\,}l r@{\,$\pm$\,}l r@{\,$\pm$\,}l r@{\,$\pm$\,}l}
\hline
\ & \mcc{\bf This work (final)} & \mc{A10 (Case 1)} & \mc{A10 (Case 2)} & \mc{A10 (Case 3)} & \mc{\citet{Triaud+10aa}} & \mc{A11} \\
\hline
$e$                    & \mcc{0.0 adopted}             & \erc{0.129}{0.106}{0.068}   & \erc{0.237}{0.068}{0.069}   & \mc{0.0 adopted}            & \mc{0.0 adopted}         & \erc{0.028}{0.015}{0.018} \\
$M_{\rm A}$    (\Msun) & 1.286    & 0.076    & 0.020   & \mc{$1.20 \pm 0.12$}        & \mc{$1.16 \pm 0.12$}        & \mc{$1.25 \pm 0.13$}        & 1.20 & 0.12               & 1.306 & 0.026   \\
$R_{\rm A}$    (\Rsun) & 1.583    & 0.040    & 0.008   & \erc{1.38}{0.20}{0.18}      & \erc{1.200}{0.081}{0.080}   & \mc{$1.566 \pm 0.073$}      & \erc{1.579}{0.067}{0.060} & 1.572 & 0.056   \\
$\log g_{\rm A}$ (cgs) & 4.149    & 0.014    & 0.002   & \mc{$4.23 \pm 0.12$}        & \mc{$4.341 \pm 0.068$}      & \erc{4.143}{0.032}{0.031}   & \mc{ }                    & 4.161 & 0.026   \\
$\rho_{\rm A}$ (\psun) & \mcc{$0.324 \pm 0.012$}       & \erc{0.45}{0.23}{0.15}      & \erc{0.67}{0.16}{13}        & \erc{0.323}{0.035}{0.028}   & \erc{0.304}{0.016}{0.020} & 0.336 & 0.030   \\[2pt]
$M_{\rm b}$    (\Mjup) & 0.477    & 0.033    & 0.005   & \erc{0.490}{0.059}{0.056}   & \erc{0.496}{0.064}{0.060}   & \erc{0.498}{0.059}{0.056}   & \erc{0.453}{0.043}{0.035} & 0.486 & 0.032   \\
$R_{\rm b}$    (\Rjup) & 1.932    & 0.052    & 0.010   & \erc{1.74}{0.26}{0.23}      & \mc{$1.51 \pm 0.10$}        & \mc{$1.97 \pm 0.10$}        & \erc{1.986}{0.089}{0.074} & 1.991 & 0.081   \\
$g_{\rm b}$     (\mss) & \mcc{$3.16 \pm 0.20$}         & \erc{3.63}{1.4}{0.9}        & \erc{5.0}{1.1}{0.9}         & \erc{2.92}{0.36}{0.33}      & \mc{ }                    &  2.81 & 0.27    \\
$\rho_{\rm b}$ (\pjup) & 0.0618   & 0.0048   & 0.0003  & \erc{0.092}{0.054}{0.032}   & \erc{0.144}{0.042}{0.031}   & \erc{0.0648}{0.0106}{0.0090}& \mc{ }                    & 0.0616 & 0.0080 \\[2pt]
\Teq\              (K) & \mcc{$1755 \pm   28$}         & \erc{1662}{113}{110}        & \mc{$1557 \pm 55$}          & \erc{1756}{26}{30}          & \mc{ }                    &  1771 & 35      \\
\safronov\             & 0.0196   & 0.0012   & 0.0001  & \mc{ }                      & \mc{ }                      & \mc{ }                      & \mc{ }                    & \mc{ }          \\
$a$               (AU) & 0.05125  & 0.00099  & 0.00027 & \erc{0.0501}{0.0017}{0.0018}& \erc{0.0494}{0.0017}{0.0018}& \erc{0.0507}{0.0017}{0.0018}& 0.0500 & 0.0017           & 0.05150&0.00034 \\
Age              (Gyr) &\ermcc{2.7}{0.6}{1.0}{0.6}{0.6}& \erc{3.0}{0.9}{2.6}         & \erc{1.2}{2.8}{1.2}         & \erc{3.1}{1.1}{0.8}         & \mc{ }                    & 2.65 & 0.25     \\
\hline \end{tabular} \end{table*}

We also modelled the Cousins $I$-band light curve from the Euler Telescope presented by A10, in order to provide a direct comparison with our results. The LD-fit/fix option was also the best, and correlated noise was found to be important for all photometric parameters. A plot of the best fit is shown in Fig.\,\ref{fig:lcfit} and the corresponding parameters are tabulated in Table\,\ref{tab:lcfit}. The agreement between the Danish and Euler data is poor, especially for $k$. The Danish results should be more reliable as they are based on three transits obtained using an equatorially-mounted telescope in excellent photometric conditions. The Euler data are more scattered, cover only one transit with a gap near midpoint, and were obtained from an alt-az telescope so suffer from continual changes in the light path over the duration of the observing sequence. We therefore adopted the Danish results as final.

Table\,\ref{tab:lcfit} also shows a comparison between our photometric parameters and those published by other researchers. The values from \citet{Triaud+10aa} and A11 are in good agreement with our results, except for the parameter $k$. This is as expected, because those authors had only the Euler light curve with which to constrain the transit shape. The parameters from A10 are more discrepant, primarily because their favoured solution is for a large orbital eccentricity.

To demonstrate the influence of eccentricity, we have rerun the above analyses on the Danish data for three eccentric orbits, applying constraints using the method described by \citet{Me+09apj}. For the first set of constraints we used $e\cos\omega = 0.036 \pm 0.033$ and $e\sin\omega = -0.10 \pm 0.13$ (A10), for the second we adopted $e\cos\omega = 0.00352 \pm 0.00075$ and $e\sin\omega = -0.027 \pm 0.019$ (A11), and for the third we specified $e\cos\omega = 0.00093 \pm 0.00068$ and $e\sin\omega = -0.027 \pm 0.019$ (Sect.\,\ref{sec:lc:e}).

The results for these three alternative sets of constraints are shown in Table\,\ref{tab:lcfit}. It can be seen that $k$ is unaffected but, as expected, $r_{\rm A}$$+$$r_{\rm b}$ and thus $r_{\rm A}$ and $r_{\rm b}$ are very dependent on the treatment of eccentricity. That large eccentricities can be ruled out is vital for precisely measuring the properties of WASP-17. The very small $e\cos\omega$ value allowed by the transit and occultation timings (Sect.\,\ref{sec:lc:e}) results in photometric parameters which are close to the 1$\sigma$ errorbars of the zero-eccentricity result, and illustrates the small change in the measured system properties to be expected if the 1.4$\sigma$ measurement of $e\cos\omega$ turns out to be real.


\begin{table*} \centering \caption{{\label{tab:bigplanets} Compilation of selected physical
properties of TEP systems containing a planet larger than 1.6\Rjup. The projected spin-orbit
misalignment measurable from the Rossiter-McLaughlin effect is denoted by $\lambda$. All
asymmetric errorbars have been averaged, and the statistical and systematic errorbars have
been added in quadrature, in order to fit into the Table.
\newline
{\bf References:}
(1) \citet{Hartman+11apj};
(2) \citet{Hartman+12};
(3) \citet{Latham+10apj};
(4) \citet{Me12mn};
(5) \citet{Fortney+11apjs};
(6) \citet{Konacki+05apj};
(7) \citet{Me10mn};
(8) \citet{Konacki+03nat};
(9) \citet{Snellen+09aa};
(10) \citet{Mandushev+07apj};
(11) \citet{Hebb+09apj};
(12) \citet{Joshi+09mn};
(13) \citet{Enoch+11aj}.
}}
\begin{tabular}{l p{0.7cm} p{0.7cm} p{0.7cm} p{0.7cm} p{1.3cm} p{0.7cm} p{0.7cm} p{0.7cm} p{0.7cm} p{1.5cm} p{1.0cm} p{1.15cm}} \hline
System      & $M_{\rm A}$ (\Msun) & $R_{\rm A}$ (\Rsun) & \Teff\ (K) & \FeH\ & $e$ & Period (d) & $M_{\rm b}$ (\Mjup) & $R_{\rm b}$ (\Rjup) & \Teq\ (K) & $\lambda$ \ ($^\circ$) & $\log R^\prime_{\rm HK}$ & Refs \\
\hline
HAT-P-32    & 1.160 $\pm$0.041 & 1.219 $\pm$0.016 & 6207 $\pm$88  & $-$0.04 $\pm$0.08 & 0.0                      & 2.150 & 0.860 $\pm$0.164 & 1.789 $\pm$0.025 & 1786 $\pm$26 &                       &          & 1         \\  
HAT-P-33    & 1.375 $\pm$0.040 & 1.637 $\pm$0.034 & 6446 $\pm$88  & $+$0.07 $\pm$0.08 & 0.0                      & 3.475 & 0.762 $\pm$0.101 & 1.686 $\pm$0.045 & 1782 $\pm$28 &                       &          & 1         \\  
HAT-P-40    & 1.512 $\pm$0.073 & 2.206 $\pm$0.061 & 6080 $\pm$100 & $+$0.22 $\pm$0.10 & 0.0                      & 4.457 & 0.615 $\pm$0.038 & 1.730 $\pm$0.062 & 1770 $\pm$33 &                       & $-$5.12  & 2 \\
HAT-P-41    & 1.418 $\pm$0.047 & 1.683 $\pm$0.047 & 6390 $\pm$100 & $+$0.21 $\pm$0.10 & 0.0                      & 2.694 & 0.800 $\pm$0.102 & 1.685 $\pm$0.064 & 1941 $\pm$38 &                       & $-$5.04  & 2 \\
Kepler-7    & 1.41  $\pm$0.10  & 2.028 $\pm$0.038 & 5933 $\pm$50  & $+$0.11 $\pm$0.05 & 0.0                      & 4.885 & 0.453 $\pm$0.068 & 1.649 $\pm$0.038 & 1619 $\pm$15 &                       & $-$5.099 & 3,4       \\  
Kepler-12   & 1.16  $\pm$0.12  & 1.490 $\pm$0.050 & 5947 $\pm$100 & $+$0.07 $\pm$0.04 & 0.0                      & 4.438 & 0.430 $\pm$0.049 & 1.695 $\pm$0.030 & 1485 $\pm$25 &                       &          & 5,4       \\  
OGLE-TR-10  & 1.277 $\pm$0.083 & 1.520 $\pm$0.100 & 6075 $\pm$86  & $+$0.28 $\pm$0.10 & 0.0                      & 3.101 & 0.68  $\pm$0.15  & 1.706 $\pm$0.054 & 1702 $\pm$54 &                       &          & 6,7       \\  
OGLE-TR-56  & 1.34  $\pm$0.10  & 1.737 $\pm$0.045 & 6119 $\pm$62  & $+$0.25 $\pm$0.08 & 0.0                      & 1.212 & 1.41  $\pm$0.18  & 1.734 $\pm$0.059 & 2482 $\pm$30 &                       &          & 8,4       \\  
OGLE-TR-L9  & 1.43  $\pm$0.10  & 1.499 $\pm$0.043 & 6933 $\pm$58  & $-$0.05 $\pm$0.20 & 0.0                      & 2.486 & 4.4   $\pm$1.5   & 1.633 $\pm$0.046 & 2034 $\pm$22 &                       &          & 9,4       \\  
TrES-4      & 1.339 $\pm$0.086 & 1.834 $\pm$0.087 & 6200 $\pm$75  & $+$0.14 $\pm$0.09 & 0.0                      & 3.554 & 0.897 $\pm$0.075 & 1.735 $\pm$0.072 & 1805 $\pm$40 & $+$6.3 $\pm$ 4.7      & $-$5.104 & 10,4      \\  
WASP-12     & 1.38  $\pm$0.19  & 1.619 $\pm$0.079 & 6300 $\pm$100 & $+$0.30 $\pm$0.10 & \er{0.018}{0.024}{0.014} & 1.091 & 1.43  $\pm$0.14  & 1.825 $\pm$0.094 & 2523 $\pm$45 &                       & $-$5.500 & 11,4      \\  
WASP-14     & 1.35  $\pm$0.12  & 1.666 $\pm$0.087 & 6475 $\pm$100 & $+$0.0  $\pm$0.2  & 0.088 $\pm$0.003         & 2.244 & 7.90  $\pm$0.47  & 1.633 $\pm$0.092 & 2090 $\pm$59 & $-$33.1 $\pm$ 7.4     & $-$4.923 & 12,4      \\  
WASP-17     & 1.286 $\pm$0.079 & 1.583 $\pm$0.041 & 6550 $\pm$100 & $-$0.25 $\pm$0.09 & 0.0                      & 3.735 & 0.477 $\pm$0.033 & 1.932 $\pm$0.053 & 1755 $\pm$28 & \er{-148.5}{4.2}{5.4} & $-$5.331 & This work \\  %
WASP-48     & 1.19  $\pm$0.05  & 1.75  $\pm$0.09  & 6000 $\pm$150 & $-$0.12 $\pm$0.12 & 0.0                      & 2.144 & 0.98  $\pm$0.09  & 1.67  $\pm$0.10  & 2030 $\pm$70 &                       &          & 13        \\  
\hline \end{tabular} \end{table*}

\section{The physical properties of WASP-17}

The physical properties of the WASP-17 system can be obtained from the adopted photometric parameters (Table\,\ref{tab:lcfit}), the orbital velocity amplitude of the star ($K_{\rm A} = 52.7 \pm 2.9$\ms; \citealt{Triaud+10aa}), its effective temperature (\Teff) and metallicity (\FeH), and a constraint from theoretical stellar evolutionary models. Full details of our approach can be found in \citet{Me09mn,Me10mn}.

One immediate problem faced here is the diversity of the published \Teff\ measurements of WASP-17\,A: A10 find $6550 \pm 100$\,K from analysis of \'echelle spectra; \citet{Triaud+10aa} obtain $6650 \pm 80$\,K from similar observations, and \citet{Maxted++11mn} deduce $6500 \pm 75$\,K from the InfraRed Flux Method \citep[IRFM;][]{BlackwellShallis77mn,Blackwell++80aa}. We adopted $\Teff = 6550 \pm 100$\,K as this encompasses all three determinations but leans towards the IRFM value, which should be the method with the least dependence on stellar theory and analysis technique. The corresponding metallicity is $\FeH = -0.25 \pm 0.09$ (A10).

Our approach was to guess a value of the orbital velocity of the planet ($K_{\rm b}$) and combine it with the photometric parameters and $K_{\rm A}$ to calculate the physical properties of both bodies using standard formulae \citep[e.g.][]{Hilditch01book}. The predicted properties of the star for the calculated mass were then found via interpolation in a set of theoretical model predictions. $K_{\rm b}$ was iteratively adjusted to find the best agreement between the known $r_{\rm A}$ and calculated $\frac{R_{\rm A}}{a}$, and between the measured and predicted \Teff\ values. This was done for ages ranging from zero to the point at which the star evolves to $\logg < 3.5$, leading to a final set of best-fitting physical properties and age for the system.

Statistical errors in the input parameters were propagated using a perturbation analysis \citep{Me++05aa}. Systematic errors arising from the use of theoretical predictions were estimated by running separate solutions for each of five independent stellar model tabulations: {\it Claret} \citep{Claret04aa}, {\it Y$^2$} \citep{Demarque+04apjs}, {\it Teramo} \citep{Pietrinferni+04apj}, {\it VRSS} \citep{Vandenberg++06apjs} and {\it DSEP} \citep{Dotter+08apjs}. Finally, a model-independent set of results was generated using an empirical calibration of stellar properties found from well-studied eclipsing binaries, a process we label ``dEB constraint''. The empirical calibration follows the approach introduced by \citet{Enoch+10aa} but with the improved calibration coefficients derived by \citet{Me11mn}.

The results for each approach are given in Table\,\ref{tab:model}. The mass, radius, surface gravity and density of the star are denoted by $M_{\rm A}$, $R_{\rm A}$, $\log g_{\rm A}$ and $\rho_{\rm A}$, and of the planet by $M_{\rm b}$, $R_{\rm b}$, $g_{\rm b}$ and $\rho_{\rm b}$. The orbital semimajor axis is $a$, \Teq\ is the equilibrium temperature of the planet (neglecting albedo and heat redistribution) and $\Theta$ is the \citet{Safronov72} number. We find that the agreement between the model sets is excellent except for the {\it DSEP} models, which are quite discrepant. \citet{Me11mn} noticed that the agreement between models deteriorates at lower metallicities, as is experienced here. Our final physical properties for the WASP-17 system (Table\,\ref{tab:fmodel}) were therefore calculated from the results obtained using the other four model sets ({\it Claret}, {\it Y$^2$}, {\it Teramo} and {\it VRSS}). Table\,\ref{tab:fmodel} also contains results from published studies of WASP-17, which are in good agreement overall but contain some optimistic errorbars.


\section{What causes such large planet radii?}

\begin{figure} \includegraphics[width=0.48\textwidth,angle=0]{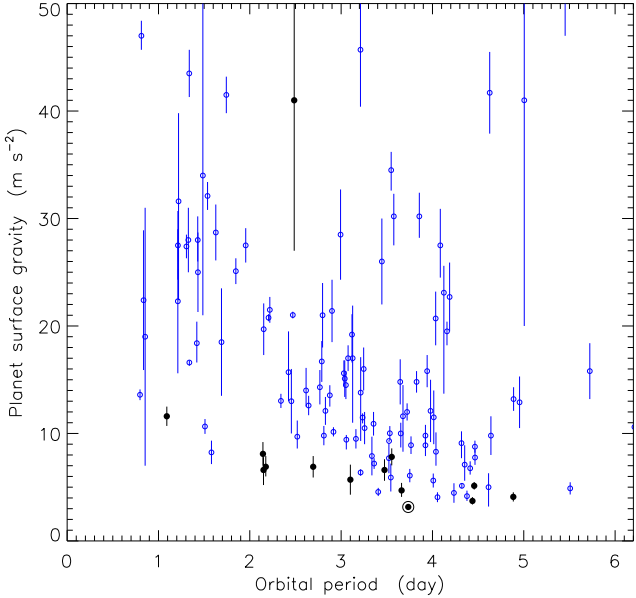}
\caption{\label{fig:porb-g2} Plot of orbital period versus surface gravity
for the known transiting planets. Planets with radii above 1.6\Rjup\ are
shown using black filled circles, and WASP-17\,b is highlighted with an
extra circle. The other planets are shown with lighter open circles.} \end{figure}

We have measured the radius of WASP-17\,b to be $R_{\rm b} = 1.932 \pm 0.053$\Rjup\ (adding the statistical and systematic errors in quadrature), confirming its status as the largest planet currently known. We also find no significant evidence for orbital eccentricity, removing an additional contribution to the uncertainties of the measured properties of the system.

Its closest competitor, HAT-P-32\,b, suffers a similar problem concerning the influence of orbital shape on the resulting planetary radius \citep{Hartman+11apj}. However, in its case the possibilities are switched because the eccentric-orbit alternative has a positive $e\sin\omega$ compared to the previously-postulated negative $e\sin\omega$ for WASP-17. The preferred solution for HAT-P-32, with a circular orbit, results in a smaller radius ($R_{\rm b} = 1.789 \pm 0.025$\Rjup) compared to the eccentric-orbit alternative ($R_{\rm b} = 2.037 \pm 0.099$\Rjup\ with $e = 0.163 \pm 0.061$).

{The other \reff{twelve} TEPs with radii above $1.6$\Rjup\ are listed in Table\,\ref{tab:bigplanets} with a summary of the physical properties of them and their parent stars. This list has been boosted by the addition of two objects, OGLE-TR-56 and WASP-14, based on major revisions to their radii by \citet{Me12mn}. \reff{The newly-discovered systems WASP-78 and WASP-79 \citep{Smalley+12} are not included here as the radii of their planets are uncertain.} We now discuss the properties of the \reff{fourteen} planets with radii above $1.6$\Rjup.}



Firstly, Fig.\,\ref{fig:porb-g2} shows that their orbital period distribution\footnote{Data taken from the Transiting Extrasolar Planet Catalogue (TEPCat, available at: {\tt http://www.astro.keele.ac.uk/jkt/tepcat/}} is not exceptional, and that they are consistent with the known correlation between period and surface gravity \citep{Me++07mn}. The masses of all but two of them (OGLE-TR-L9 and WASP-12) are in the interval 0.4--1.0\Mjup. These objects also do not represent a high-eccentricity population: all have orbits which are (or are almost) circular\footnote{It has been noticed that orbital eccentricity is correlated with planet mass \citep{Me+09apj} but plots of planet radius versus eccentricity (not shown here) indicate that there is -- if anything -- a deficit of large-radius eccentric planets.}. Tidal heating \citep{Bodenheimer++01apj,Jackson++08apj,IbguiBurrows09apj} is therefore not a viable proposition to explain their large radii. {Orbital misalignment may be relevant: WASP-14 is misaligned \citep{Johnson+09pasp}, WASP-17 is retrograde \citep{Triaud+10aa}, and TrES-4 is axially aligned \citep{Narita+10pasj2}. Thus two of the three planets with obliquity measurements are misaligned.}

\begin{figure} \includegraphics[width=0.48\textwidth,angle=0]{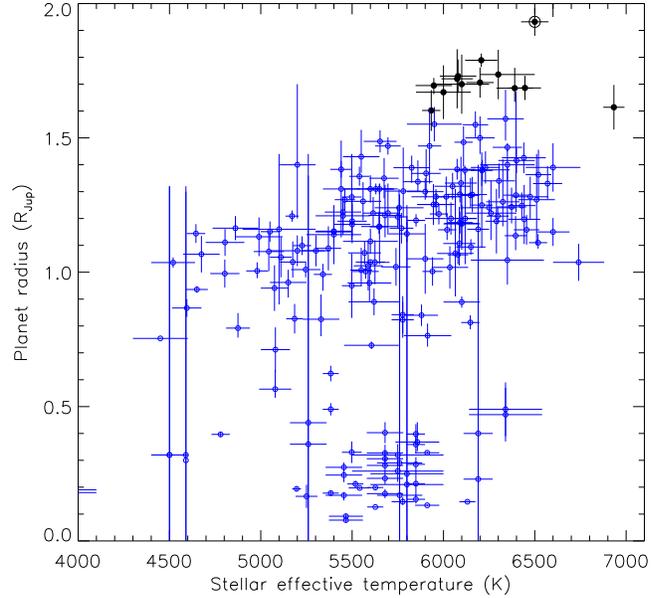}
\caption{\label{fig:teff-r2} Plot of stellar \Teff\ versus planet radius
for the known transiting systems. Other comments are the same as in
Fig.\ref{fig:porb-g2}.} \end{figure}

\begin{figure} \includegraphics[width=0.48\textwidth,angle=0]{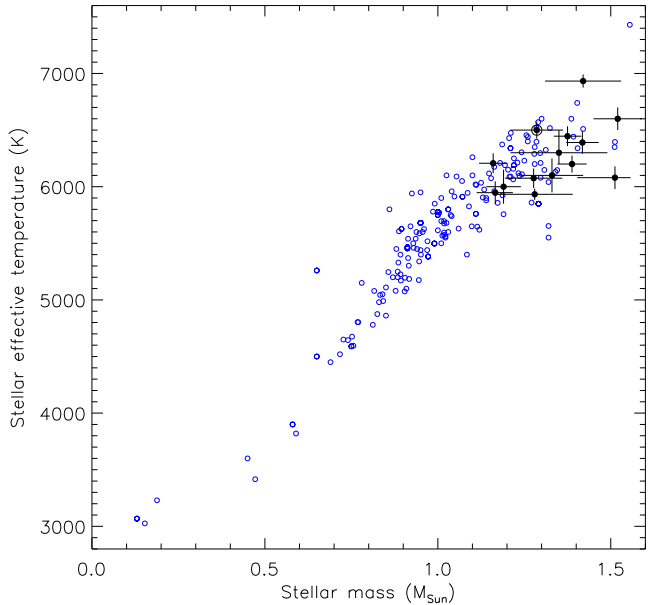}
\caption{\label{fig:m1-teff} Plot of stellar mass versus effective temperature
for the known transiting planets (data taken from TEPCat). Other comments are the
same as in Fig.\ref{fig:porb-g2}. The errorbars of the planets have been surpressed
for clarity, with the exception of those \reff{fourteen} with a large radius.} \end{figure}

However, a noticable feature of the \reff{fourteen} large planets is that they orbit host stars with $\Teff > 5900$\,K and $M_{\rm A} > 1.15$\Msun. This establishes a connection between a bloated planet and a hot star. Fig.\,\ref{fig:teff-r2} and Fig.\,\ref{fig:m1-teff} show that the large planets are preferentially associated with hotter and more massive host stars. The association does not work the other way: such stars also possess TEPs with smaller radii representative of the general planet population. A correlation with host star \FeH\ was suspected but not found. An important factor in bloating planets above their expected size therefore seems to be the \Teff\ of their host star. This may be due to the enhanced UV flux from such stars, but such a possibility does not explain why less inflated planets are found around stars with $\Teff > 5900$\,K.

A similar situation occurs with irradiation: the large planets have high \Teq s (or equivalently large specific incident stellar fluxes) of $1600$\,K or more, but so do many other smaller planets. The connection between inflated radii and high \Teq\ is well-known \citep[see][and references therein]{Baraffe++10rpph,Enoch+12aa} but the simultaneous existence of small planets with high \Teq\ is not yet understood. \citet{Laughlin++11apj} quantified the correlation between the radius anomaly (observed radius versus those predicted by the models of \citealt{Bodenheimer++03apj}) of TEPs and their \Teq s, in the form of a power law. Whilst their fig.\,2 exhibits appreciable evidence for this claim, the large planets discussed here remain outliers even in that diagram.

We have searched for values for the chromospheric activity indicator $\log R^{\prime}_{\rm HK}$ \citep{Noyes+84apj} for the host stars of our large planets. \citet{Knutson++10apj} give values of $-5.331$ for WASP-17, $-5.099$ for Kepler-7, $-5.104$ for TrES-4, $-5.500$ for WASP-12 {and $-4.923$ for WASP-14}. These $\log R^{\prime}_{\rm HK}$ values are representative of inactive stars, suggesting a correlation between inflated planetary radii and low chromospheric activity. HAT-P-32 and HAT-P-33 possess values of the related $S$ index \citep{Vaughan++78pasp} which indicate a low activity level, despite their comparatively high rotation rates and velocity jitter \citep{Hartman+11apj}. \reff{HAT-P-40 and HAT-P-41 have similarly quiet $\log R^{\prime}_{\rm HK}$ values of $-5.12$ and $-5.04$ \citep{Hartman+12}.}

Therefore \reff{eight} of the \reff{fourteen} stars in question have measured activity indicators, and all suggest low chromospheric activity. Perhaps the increased high-energy photon flux from more active stars acts against large planetary radii. \citet{Knutson++10apj} found a correlation with the atmospheric properties of planets, in that inactive stars possess planets with temperature inversions whereas planets around active stars do not have inversions. However, \citet{Hartman10apj} found no correlation between $\log R^{\prime}_{\rm HK}$ and planet radius, so the low values for the \reff{fourteen} host stars in question may be an artefact of their \Teff\ distribution.

Another possibility to explain the large planetary radii is that the more massive stars, around which the larger planets are found, have shorter main-sequence lifetimes. They will therefore be on average younger than the less massive TEP host stars. The large planets could simply be at an earlier stage of their evolution. This is, however, at odds with the low activity levels of the host stars, which implies that they are not particularly young. The models by \citet[][their fig.\,5]{Fortney++07apj} show that it is possible for planets to be 2.0\Rjup\ or above if they are young (of order $10^7$\,yr) and low-mass (less than 1\Mjup). These criteria are not satisfied by any of our large planets. Explaining the radii of the large TEPs is therefore only viable if there is a large systematic error in our estimation of the age of their host stars, which is unlikely but certainly not impossible.

\subsection{Can observational biases explain the properties of large planets?}

\begin{figure} \includegraphics[width=0.48\textwidth,angle=0]{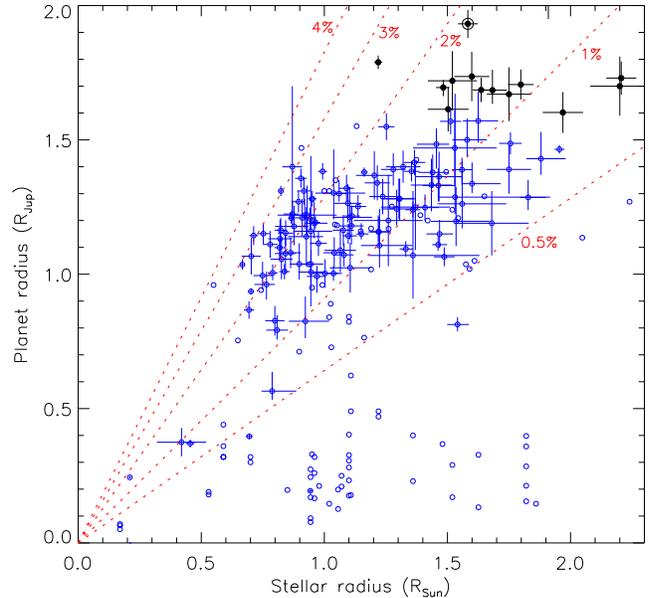}
\caption{\label{fig:r1-r2} Plot of stellar versus planet radius for the
known TEP systems. Other comments are the same as in Fig.\ref{fig:porb-g2}.
Points without errorbars represent those discovered via the space telescopes
CoRoT and {\it Kepler}, whose discovery biases are very different from
those of ground-based surveys.} \end{figure}

An important question is whether the correlation between large planets and hot and massive host stars is real, or is it merely a phantom arising from observational selection effects? The correlation could easily be suspected to arise from detection biases as a function of transit depth. Fig.\,\ref{fig:r1-r2} shows a plot of the radii of the host stars versus their planets. We have overlaid dotted lines to indicate loci of approximately constant transit depth, calculated using {\sc jktebop} and assuming quadratic LD with both coefficients equal to $0.30$. The subset of planets discovered via space telescopes are down-played in Fig.\,\ref{fig:r1-r2}, as these should be much less biased against finding systems with shallow transits. It is immediately clear that the large planets do not stand out in this diagram as having unusually deep, and therefore easily detectable, transits.

Fig.\,\ref{fig:r1-r2} shows that most known TEPs have transit depths of about 1--3\%. Firstly, we are straightforwardly biased against finding small planets around hotter (and therefore bigger) stars, as the transit depths in these systems are small. The relative paucity of small planets around big stars is plausibly explained by lower detection probabilities for transits less than 0.5\% deep. Secondly, the absence of large planets around small stars might be explicable by either natural rarity or a bias against deep transits. Such objects may have a low follow-up priority within planet-search consortia if it is believed that deep transits are associated with false positives such as eclipsing binary star systems. As an example, if WASP-17\,b orbited \reff{an unevolved} 0.9\Msun\ star then the transit depth would be roughly 5\%. 

None of the observables discussed above pre-determines the existence of big planets. They are found only around hotter stars, but some such stars possess small planets. They have quite high \Teq s, but so do many smaller planets. They do not give rise to unusually deep or shallow transits, which rules out the more simple observational biases, and the metallicities of their host stars are not exceptional. The question of what causes their bloated radii remains unsettled.


\section{Summary and conclusions}

Whilst WASP-17 was widely regarded to be the largest known planet, its radius was uncertain as it was based primarily on one follow-up transit light curve which shows moderate correlated noise. In this work we present three new high-precision transit light curves, obtained using telescope-defocussing techniques, and improve the measurement of its radius.

We have refined the orbital ephemeris of the system using our new data, which increase the temporal baseline by 3.1\,years. Our revised orbital period is {$4.0 \pm 0.6$\,s longer than previous measurements, a difference of 6.6$\sigma$}, and this change is sufficient to bring the observed time of occultation (A11) into line with that expected for a circular orbit. Further observations would allow this result to be checked. In the case of WASP-17, circularity of the orbit favours a larger value for the planetary radius.

We have modelled our new data, along with the single follow-up light curve presented in the discovery paper (A10), using the {\sc jktebop} code. We paid careful attention to the treatment of limb darkening and to obtaining reliable uncertainties. The physical properties of the system were then derived using our new photometric and published spectroscopic results. Remaining uncertainties and discrepancies in the existing \Teff\ and \FeH\ measurements of the host star impose a bottleneck on the quality of the resulting properties, and new radial velocity observations would also be useful in refining the measurement of the planet's mass and therefore its density.

WASP-17\,b is the largest known planet, with a radius of $R_{\rm b} = 1.932 \pm 0.053$\Rjup. Another {eleven} planets are known with radii greater than 1.6\Rjup. They are found only around comparatively hot ($\Teff > 5900$\,K) and massive ($M_{\rm A} > 1.15$\Msun) stars, and have correspondingly high equilibrium temperatures ($\Teq > 1600$\,K with the exception of Kepler-12) and equivalently incident fluxes. But other stars of similar mass and \Teff\ possess smaller planets, whilst other planets with similar \Teq\ (or equivalently specific incident flux) do not have such enlarged radii. One possible discriminating feature is that all \reff{eight} of the \reff{fourteen} host stars with measured activity indicators are chromospherically inactive.

The set of \reff{fourteen} large planets do not have unusual transit depths. However, planets of this size around cooler stars may have an anomalously low discovery rate if their deep transits (of order 5\%) discount them from detailed follow-up observations. High-precision radial velocity measurements are expensive in terms of telescope time, so dubious TEP candidates with unexpectedly deep transits may be prematurely rejected as false positives. A re-evaluation of such objects will either yield scientifically valuable discoveries, or dismiss the existence of large planets around small stars.

The \reff{fourteen} planets with radii greater than 1.6\Rjup\ all have circular (or nearly circular) orbits, so their large size cannot easily be attributed to tidal heating. Of the {three} published measurements of the axial alignments of these planets, one reveals a retrograde orbit, {one a misaligned orbit, and one} indicates alignment. Observations of the Rossiter-McLaughlin effect on the remaining \reff{eleven} systems could either verify or discount the possibility that axial alignment is a relevant aspect of large planets.


\section*{Acknowledgments}

%
%
%
%
%
%
%

The reduced light curves presented in this work will be made available at the CDS ({\tt http://cdsweb.u-strasbg.fr/}) and at {\tt http://www.astro.keele.ac.uk/$\sim$jkt/}. J\,Southworth acknowledges financial support from STFC in the form of an Advanced Fellowship. The operation of the Danish 1.54m telescope is financed by a grant to UGJ from the Danish Natural Science Research Council. We also acknowledge support from the European Community's Seventh Framework Programme (FP7/2007-2013/) under grant agreement Nos.\ 229517 and 268421, and support from the ASTERISK project (ASTERoseismic Investigations with SONG and Kepler) funded by the European Research Council (grant agreement No.\ 267864) and from the Centres of Excellence Centre for Star and Planet Formation (StarPlan) and Stellar Astrophysics Centre (SAC) funded by The Danish National Research Foundation. MD, MH, CL and CS are thankful to the Qatar National Research Fund (QNRF), a member of Qatar
Foundation, for support by grant NPRP 09-476-1-078. SG and XF acknowledge the support from NSFC under the grant No.\ 10873031. DR (boursier FRIA), OW (FNRS research fellow) and J\,Surdej acknowledge support from the Communaut\'e fran\c{c}aise de Belgique -- Actions de recherche concert\'ees -- Acad\'emie Wallonie-Europe.

\bibliographystyle{mn_new}

\end{document}